# Optimizing Continued Fraction Expansion Based IIR Realization of Fractional Order Differ-Integrators with Genetic Algorithm

Saptarshi Das, Basudev Majumder, Anindya Pakhira, Indranil Pan, Shantanu Das, Amitava Gupta

*Abstract*—Rational approximation of fractional order (FO) differ-integrators via Continued Fraction Expansion (CFE) is a well known technique. In this paper, the nominal structures of various generating functions are optimized using Genetic Algorithm (GA) to minimize the deviation in magnitude and phase response between the original FO element and the rationalized discrete time filter in Infinite Impulse Response (IIR) structure. The optimized filter based realizations show better approximation of the FO elements in comparison with the existing methods and is demonstrated by the frequency response of the IIR filters.

## I. INTRODUCTION

FRACTIONAL order systems have gained wide attention in recent years from different research communities due to their added flexibility and improved performance over their integer order counterparts in a wide variety of fields ranging from control, signal processing to biological applications [1]. Recent hardware implementation of fractional order elements [2] have given more impetus to the implementation aspect of these systems by means of proposing various forms of realizations and approximations which mimic the original system to a certain degree of accuracy and at the same time can be easily implemented in real hardware with the help of simple mathematical operators [3].

For practical purposes, a band limited implementation of the FO elements is important. This indicates that FO elements which are basically infinite dimensional linear filters needs to be approximated with finite dimensional transfer functions in a specified band of frequencies of practical interest [4]-[5]. There are mainly two methods of discretization viz. indirect and direct method [6]. The indirect discretization method is accomplished in two steps. Firstly the frequency domain fitting is done in continuous time domain and then the fitted continuous time transfer function is discretized. Direct discretization based methods [7] include the application of Power Series Expansion (PSE), Continuous Fractional Expansion (CFE) [8], MacLaurin Series Expansion [9] etc with a suitable generating function. The mapping relation or formula for conversion from continuous time to discrete time operator ($s \leftrightarrow z$) is known as the generating function. Among the family of expansion methods, CFE based digital realization has been extensively studied with various types of generating functions like Tustin [10], Simpson [11], Al-Alaoui [12], mixed Tustin-Simpson [13], mixed Euler-Tustin-Simpson [14], impulse response based [15] and other higher order generating functions [16]-[18].

This paper focuses on the CFE based realization of the fractional order differ-integrators with an optimization based approach for the mixed type generating functions. In particular the weights of various composite generating functions like Al-Alaoui type [19], Chen-Vinagre type [13] etc. are optimized with a stochastic evolutionary algorithm known as Genetic Algorithm (GA) to minimize the discrepancies between the magnitude and phase response of the original FO differ-integrator and the high order IIR filter representing its band-limited discrete time realization.

The rest of the paper is organized as follows. Section II builds up the theoretical framework for the present proposition and discusses the IIR filter realization problem in the light of the optimization framework. Section III illustrates the simulation results and highlights the major findings with appropriate discussions. The paper ends in Section IV with the conclusions followed by the references.

## II. THEORETICAL FORMULATION

### A. Family of Generating Functions and Basic Concepts

Chen, Petras & Xue [7] and Chen, Vinagre & Podlubny [8] have introduced four classes of generating functions representing the discrete time rational approximation of a simple continuous-time differentiator ($s \approx H(z^{-1})$) as:

$$H_{Euler}(z^{-1}) = \left[\frac{1-z^{-1}}{T}\right] \qquad (1)$$

Here, $T$ represents the sampling time and $z$ denotes the discrete time complex frequency. It is clear that the Euler's discretization formula (1) is an extension of the backward difference technique of numerical differentiation.

$$H_{Tustin}(z^{-1}) = \left[\frac{2}{T} \cdot \frac{1-z^{-1}}{1+z^{-1}}\right] \qquad (2)$$

Manuscript received April 15, 2011. This work has been supported by the Department of Science & Technology (DST), Govt. of India under the PURSE programme.

S. Das, B. Majumder are with School of Nuclear Studies and Applications, Jadavpur University, Salt-Lake Campus, LB-8, Sector 3, Kolkata-700098, India. (E-mail: saptarshi@pe.jusl.ac.in).
A. Pakhira is with Department of Instrumentation & Electronics Engineering, Jadavpur University, Salt-Lake Campus, LB-8, Sector 3, Kolkata-700098, India.
I. Pan and A. Gupta are with Department of Power Engineering, Jadavpur University, Salt-Lake Campus, LB-8, Sector 3, Kolkata-700098, India.
Sh. Das is with Reactor Control Division, Bhabha Atomic Research Centre, Mumbai-400085, India.

The Tustin's discretization can be obtained from the basic ($s \leftrightarrow z$) mapping relation by expanding the exponential terms with their first order approximations. i.e.

$$z = e^{sT} = \frac{e^{sT/2}}{e^{-sT/2}} = \left(1 + \frac{sT}{2}\right) \Big/ \left(1 - \frac{sT}{2}\right) = \frac{2 + sT}{2 - sT} \quad (3)$$

$$\Rightarrow s = \frac{2}{T} \cdot \left(\frac{z-1}{z+1}\right)$$

Also, the well known Simpson's numerical integration formula is given by (in time domain):

$$y(n) = \frac{T}{3}[x(n) + 4x(n-1) + x(n-2)] + y(n-2) \quad (4)$$

By applying z transform on (4) it is found:

$$\frac{Y(z)}{X(z)} = H(z) = \frac{T}{3}\left(\frac{1 + 4z^{-1} + z^{-2}}{1 - z^{-2}}\right)$$

The above relationship represents a digital integrator and can be inverted to obtain a digital differentiator as:

$$H_{Simpson}(z^{-1}) = \left[\frac{3}{T} \cdot \frac{(1+z^{-1})(1-z^{-1})}{1 + 4z^{-1} + z^{-2}}\right] \quad (5)$$

Al-Alaoui has shown in [19] that the discretization formula can be improved by interpolating the classical Euler and Tustin's formula as follows:

$$H_{Al-Alaoui(\alpha)}(z^{-1}) = \alpha H_{Euler}(z^{-1}) + (1-\alpha) H_{Tustin}(z^{-1})$$

$$= \frac{\alpha T}{(1-z^{-1})} + \frac{(1-\alpha)T}{2}\left(\frac{1+z^{-1}}{1-z^{-1}}\right) = \frac{T[(1+\alpha) + (1-\alpha)z^{-1}]}{2(1-z^{-1})}$$

$$= \frac{T(1+\alpha)}{2} \cdot \frac{\left(1 + \frac{(1-\alpha)}{(1+\alpha)}z^{-1}\right)}{(1-z^{-1})} \quad (6)$$

where, $\alpha \in (0,1)$ is a user-specified weight that balances the impact of the two generating function i.e. Euler (rectangular) and Tustin (Trapezoidal) and their corresponding accuracies introduced in the discretization. Replacing $\alpha = 3/4$ in (6) produces the conventional Al-Alaoui operator as:

$$H_{Al-Alaoui(3/4)}(z^{-1}) = \frac{7T}{8} \cdot \frac{(1 + z^{-1}/7)}{(1-z^{-1})} \quad (7)$$

Generalized Al-Alaoui operator (6) shows that the IIR filter has a pole at $z = 1$ and zero varies between $z \in [-1, 0]$ for $\alpha \in [0,1]$. Thus, the operator (6) can be directly inverted to produce a stable IIR realization for a differentiator also.

Simpson type generating function (6) considers a second order polynomial fitting between two points in discrete time. The above statement can be explained better by conventional Simpson's numerical integration technique. Early discretization techniques developed by Euler and Tustin are mainly based on the First order polynomial fitting. Simpson's advancement in the discretization technique shows that one can fit higher order polynomial to obtain better accuracy. But this is not a wise technique, since expansion with higher order generating function would increase the overall order of the discrete time filter. Also, as the order of generating function increases, the region of performance in the frequency domain gets shrinked and also order of the IIR filter will be high. So we have restricted this up to second order realization only unlike [17]-[18] and optimized within a chosen structure like Al-Alaoui [19] to obtain an optimum generating function. The generalized Al-Alaoui type generating function (6) is ideal for applications where the requirement is to maximize accuracy without going for a higher order realization. The motivation behind optimum interpolation of two different discretization methods is the fact that the frequency response of a continuous time integrator lies between the Tustin and Euler/Simpson's approximation [9].

Chen & Vinagre [13] proposed a hybrid generating function that interpolates the accuracies of Simpson and Tustin's method as follows:

$$H_{Chen-Vinagre(\alpha)}(z^{-1}) = \alpha H_{Simpson}(z^{-1}) + (1-\alpha) H_{Tustin}(z^{-1})$$

$$= \frac{\alpha T}{3} \cdot \frac{(1 + 4z^{-1} + z^{-2})}{(1+z^{-1})(1-z^{-1})} + \frac{(1-\alpha)T}{2}\left(\frac{1+z^{-1}}{1-z^{-1}}\right)$$

$$= \frac{T(3-\alpha)}{6} \cdot \frac{\left(1 + (2(3+\alpha)/(3-\alpha))z^{-1} + z^{-2}\right)}{(1+z^{-1})(1-z^{-1})} \quad (8)$$

The generalized Chen-Vinagre operator (8) has two real poles at $z = \pm 1$. Its two zeros lie at $z = -1$ for $\alpha = 0$. But for any non-zero value of $\alpha \in (0,1]$ the IIR filter (8) will have non-minimum phase zero which will lead to unstable poles if the formula is inverted to represent a differentiator. Handling such unstable poles by reflecting it within the unit-circle has been extensively studied in [13].

### B. IIR Type Realization of FO Elements via Continued Fraction Expansion

Let us now, consider a fractional order integrator

$$G(s) = 1/s^\gamma, \gamma \in [0,1] \subseteq \mathbb{R}_+ \quad (9)$$

Then, with an Al-Alaoui type generating function the discrete-time realization of the FO-integrator (9) becomes

$$G(z^{-1}) = \left[\frac{T(1+\alpha)}{2} \cdot \frac{\left(1 + ((1-\alpha)/(1+\alpha))z^{-1}\right)}{(1-z^{-1})}\right]^\gamma$$

$$= \left(\frac{T(1+\alpha)}{2}\right)^\gamma \cdot CFE\left\{\left(\frac{1 + ((1-\alpha)/(1+\alpha))z^{-1}}{(1-z^{-1})}\right)^\gamma\right\} \quad (10)$$

where, $CFE\{\cdot\}$ denotes the continued fraction expansion of an irrational function $G(z)$ and is expressed as:

$$G(z) \simeq a_0(z) + \cfrac{b_1(z)}{a_1(z) + \cfrac{b_2(z)}{a_2(z) + \cfrac{b_2(z)}{a_3(z) + \cdots}}} \quad (11)$$

$$= a_0(z) + \frac{b_1(z)}{a_1(z)} + \frac{b_2(z)}{a_2(z)} + \frac{b_3(z)}{a_3(z)} + \cdots$$

In this paper, to approximate the FO integrator (9), the CFE in (10) is carried out using the symbolic computation capabilities of the Maple Toolbox for MATLAB. Maple is widely used technical computing software used by Engineers and Scientists worldwide, with an advanced symbolic computation engine. The Maple Toolbox for MATLAB tightly integrates with Maple, providing all the features of the Maple engine to MATLAB users. In this work, the symbolic computation capabilities of Maple have been used within the MATLAB environment for evaluating the CFE of the operator in question and manipulating the obtained expressions to a form suitable for the native MATLAB functions to work with. The present work is an extension of the MATLAB routines presented in [7]-[8], [12]-[13] and detailed below with few modifications.

The 'cfrac' function from the Number Theory package of Maple has been called to symbolically calculate the CFE, taking $z^{-1}$ as variable $x$. Then the convergent polynomial has been generated using the 'nthconver' function from the same package. The terms have been collected and the requisite values substituted. The numerator and denominator have been separated using 'numden' function from the MTM package. The resulting numerator and denominator were in the symbolic type. In order to use them with other native MATLAB functions, they have been converted to polynomials of type double by calling 'sym2poly' and 'double' functions from the same package in sequence.

Now, the MATLAB generates polynomial vectors with decreasing order. But here, $x = z^{-1}$. So, in order to obtain the standard filter transfer function form the numerator and denominator matrices have been flipped with the 'fliplr' MATLAB function. This transfer function represents the IIR realization of a FO integrator having a user specified order. Using the above mentioned technique few numerical studies are made for the digital realization of a semi-differentiator.

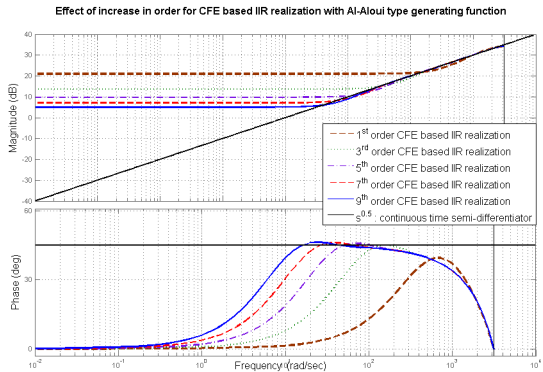

Fig. 1. Accuracies of different order of IIR realization with Al-Alaoui generating function.

Fig. 1 shows that with Al-Alaoui's generating function (7), the accuracy of the IIR realization increases especially in the phase curve for an increase in the order of realization. Also, with fixed order IIR realization with the generalized Al-Alaoui type generating function (6), the increase in the weight $\alpha$, balancing the influence of Euler and Tustin discretization formulae, better accuracy is achieved in the low frequency regimes whereas errors in the high frequency regimes also increases simultaneously (Fig. 2). This motivates us to choose $\alpha$ optimally such that it gives minimum discrepancy in both the gain and phase responses for a particular fractional order element ($s^{\pm \gamma}$) within a chosen frequency band. Fig. 3 shows the effect of decrease in the sampling time for fixed order of IIR filter realization which indicates only shift of the constant phase region towards higher frequencies. In present simulation study, the sampling time has been chosen as $T = 0.001$, as studied by Chen et al. [8].

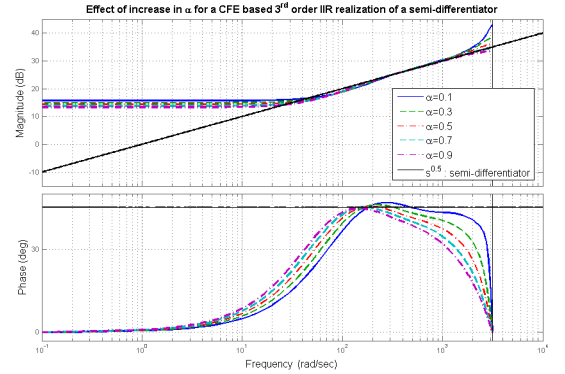

Fig. 2. Accuracies of different weights (α) of Al-Aloui type generating function interpolation for 3rd order IIR realization.

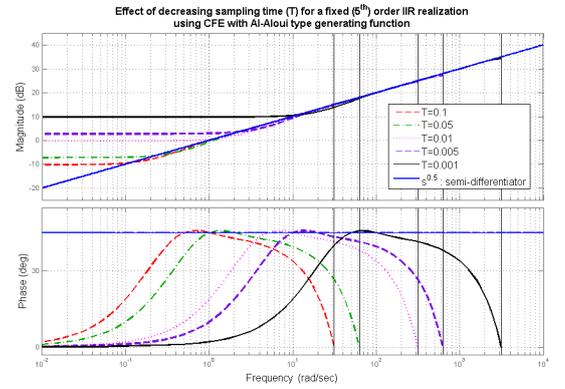

Fig. 3. Effect of change in sampling time (T) for 5th order IIR realization with Al-Alaoui type generating function.

C. *IIR Filter Realization within Optimization Framework*

The IIR filter realization is done by minimizing the weighted sum of the discrepancies between gain and phase responses of the continuous time FO element $G(s)$ and its discrete time IIR realization $\widehat{G}(z^{-1})$. The objective function ($J$) for optimum IIR realization of the FO elements is given by (12) and is minimized using GA to produce optimum value of the weight $\alpha$ for the generalized Al-Alaoui (6) and Chen-Vinagre (8) type generating functions.

$$\begin{aligned} J_{mag} &= \left\| Mag[G(s)] - Mag[\widehat{G}(z^{-1})] \right\| \\ J_{phase} &= \left\| Arg[G(s)] - Arg[\widehat{G}(z^{-1})] \right\| \\ J &= w \cdot J_{mag} + (1-w) \cdot J_{phase} \end{aligned} \quad (12)$$

The two components of the objective function (12), indicating the deviation in the magnitude and phase response of the FO differ-integrator and its digital IIR realization is evaluated within a chosen frequency band $\omega \in [10^{-4}, \omega_N]$ with $\omega_N$ being the Nyquist frequency. From Fig. 1-3 it has been already shown that with Al-Alaoui's generating function having fixed weight $\alpha = 3/4$, it is hard to maintain constant phase for a wide range of frequencies, though the gain curves are almost closer to that of the FO element. Hence, GA based IIR filter optimization has been carried out for various levels of weights $w$, balancing the discrepancies in the magnitude and phase of the realization with a chosen generating function. Also, it is to be noted that GA and other evolutionary algorithms have been extensively used in recent literatures [20]-[25] for similar digital filter optimization tasks, as also in this case.

Genetic Algorithm is a computational stochastic method for optimization based on the natural Darwinian evolution. In GA each solution vector (chromosome) is represented by bit strings which are the essentially an encoded form of the solution variables. These chromosomes evolve over successive generations through evolutionary operations like reproduction, crossover and mutation. Each set of solution vector in the mating pool is assigned a relative fitness value based on the evaluation of an objective function. The fitter individuals have a greater probability of passing on to the next generation. Newer individuals are created on probabilistic decisions from parent genes by the process of crossover. Mutation is applied at randomly selected positions of the parent gene to produce newer individuals. With these operators newer individuals are produced and the solution is iteratively refined until the objective function is minimized below a certain tolerance level or the maximum number of iterations are exceeded.

## III. SIMULATION STUDIES

### A. Optimization of Al-Alaoui Type Generating Function

Al-Alaoui [19] proposed a new class of digital integrators which combines the merits of conventional Euler and Tustin type discretization methods. In this paper, Al-Alaoui's IIR structure, representing a digital integrator is enhanced with evolutionary algorithm based optimization techniques while minimizing the objective function (12). Al-Alaoui type generating functions give slightly better accuracy when employed with GA based 5th order realization as shown in Fig. 4 and 5.

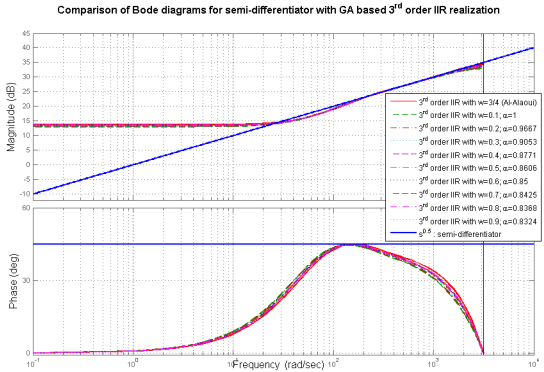

Fig. 4. Accuracies of GA based 3rd order IIR realization with various levels of gain-phase error balancing.

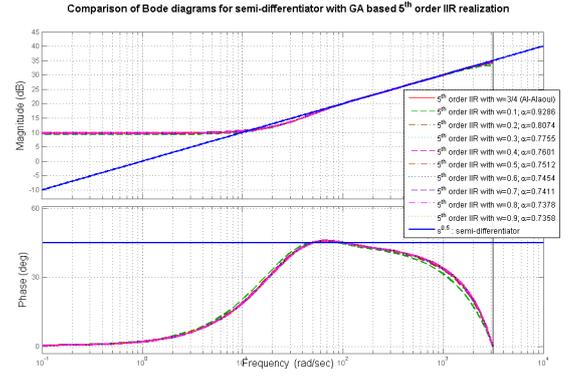

Fig. 5. Accuracies of GA based 5th order IIR realization of Al-Alaoui operator with various levels of gain-phase error balancing.

The corresponding optimization results for a semi-differentiator are reported in Table I, with optimum weights ($\alpha_{opt}$) of the generating function. It is also shown that the optimized values of the objective function ($J_{min}$) is lesser than that with the nominal Al-Alaoui operator ($J_{Al-Alaoui}$) with $\alpha = 3/4$ for each chosen value of $w$.

TABLE I
OPTIMIZATION PERFORMANCE WITH AL-ALAOUI TYPE GENERATING FUNCTION FOR 3RD AND 5TH ORDER IIR REALIZATION

| weight ($w$) in the $J$ | 3rd order IIR | | | 5th order IIR | | |
|---|---|---|---|---|---|---|
| | $J_{min}$ | $\alpha_{opt}$ | $J_{Al\text{-}Alaoui}$ ($\alpha$=0.75) | $J_{min}$ | $\alpha_{opt}$ | $J_{Al\text{-}Alaoui}$ ($\alpha$=0.75) |
| 0.1 | 770.1836 | 1.0 | 770.99 | 738.7006 | 0.9286 | 739.006 |
| 0.2 | 693.6743 | 0.9667 | 694.27 | 662.3276 | 0.8074 | 662.4116 |
| 0.3 | 617.0517 | 0.9053 | 617.55 | 585.7912 | 0.7755 | 585.8171 |
| 0.4 | 540.372 | 0.8771 | 540.84 | 509.2171 | 0.7601 | 509.2227 |
| 0.5 | 463.6698 | 0.8606 | 464.12 | 432.6281 | 0.7512 | 432.6282 |
| 0.6 | 386.9565 | 0.85 | 387.4 | 356.0319 | 0.7454 | 356.0338 |
| 0.7 | 310.2368 | 0.8425 | 310.69 | 279.4314 | 0.7411 | 279.4393 |
| 0.8 | 233.5131 | 0.8368 | 233.97 | 202.8284 | 0.7378 | 202.8449 |
| 0.9 | 156.7868 | 0.8324 | 157.26 | 126.2236 | 0.7358 | 126.2504 |

The 3rd order optimized IIR filters are reported in (13) for increasing value of the weight $w$ balancing the relative gain-phase discrepancies of a semi-differentiator:

$$G_{3,w=0.1}(z) = \frac{3795 - 6641z^{-1} + 3320z^{-2} - 415z^{-3}}{120 - 150z^{-1} + 45z^{-2} - 1.875z^{-3}}$$

$$G_{3,w=0.2}(z) = \frac{3827 - 6616z^{-1} + 3235z^{-2} - 383.5z^{-3}}{120 - 146.4z^{-1} + 41.47z^{-2} - 1.227z^{-3}}$$

$$G_{3,w=0.3}(z) = \frac{3888 - 6562z^{-1} + 3067z^{-2} - 322.7z^{-3}}{120 - 139.6z^{-1} + 34.82z^{-2} - 0.07844z^{-3}}$$

$$G_{3,w=0.4}(z) = \frac{3917 - 6534z^{-1} + 2985z^{-2} - 293.7z^{-3}}{120 - 136.3z^{-1} + 31.7z^{-2} + 0.4254z^{-3}}$$

$$G_{3,w=0.5}(z) = \frac{3934 - 6517z^{-1} + 2935z^{-2} - 276.3z^{-3}}{120 - 134.3z^{-1} + 29.86z^{-2} + 0.7122z^{-3}}$$

$$G_{3,w=0.6}(z) = \frac{3946 - 6505z^{-1} + 2902z^{-2} - 265.1z^{-3}}{120 - 133z^{-1} + 28.66z^{-2} + 0.8931z^{-3}}$$

$$G_{3,w=0.7}(z) = \frac{3954 - 6496z^{-1} + 2879z^{-2} - 257z^{-3}}{120 - 132z^{-1} + 27.82z^{-2} + 1.02z^{-3}}$$

$$G_{3,w=0.8}(z) = \frac{3960 - 6490z^{-1} + 2861z^{-2} - 250.9z^{-3}}{120 - 131.3z^{-1} + 27.17z^{-2} + 1.115z^{-3}}$$

$$G_{3,w=0.9}(z) = \frac{3964 - 6485z^{-1} + 2847z^{-2} - 246.1z^{-3}}{120 - 130.8z^{-1} + 26.67z^{-2} + 1.187z^{-3}} \quad (13)$$

Similar results can also be obtained for 5$^{th}$ order IIR realization for a semi-differentiator with different $w$ as:

$$G_{5,w=0.1} = \begin{pmatrix} 9.738\times10^5 - 2.597\times10^6 z^{-1} + 2.482\times10^6 z^{-2} \\ -1.009\times10^6 z^{-3} + 1.565\times10^5 z^{-4} - 5238z^{-5} \\ \hline 30240 - 6.496\times10^4 z^{-1} + 4.688\times10^4 z^{-2} \\ -1.269\times10^4 z^{-3} + 903.6z^{-4} + 17.93z^{-5} \end{pmatrix}$$

$$G_{5,w=0.2} = \begin{pmatrix} 1.006\times10^6 - 2.525\times10^4 z^{-1} + 2.197\times10^6 z^{-2} \\ -7.534\times10^5 z^{-3} + 7.577\times10^4 z^{-4} + 1698z^{-5} \\ \hline 30240 - 5.918\times10^4 z^{-1} + 3.614\times10^4 z^{-2} \\ -6456z^{-3} - 256.4z^{-4} + 48.77z^{-5} \end{pmatrix}$$

$$G_{5,w=0.3} = \begin{pmatrix} 1.015\times10^6 - 2.502\times10^6 z^{-1} + 2.114\times10^6 z^{-2} \\ -6.83\times10^5 z^{-3} + 5.544\times10^4 z^{-4} + 3102z^{-5} \\ \hline 30240 - 5.752\times10^4 z^{-1} + 3.322\times10^4 z^{-2} \\ -4907z^{-3} - 485.2z^{-4} + 46.38z^{-5} \end{pmatrix}$$

$$G_{5,w=0.4} = \begin{pmatrix} 1.019\times10^6 - 2.491\times10^6 z^{-1} + 2.072\times10^6 z^{-2} \\ -6.486\times10^5 z^{-3} + 4.583\times10^4 z^{-4} + 3706z^{-5} \\ \hline 30240 - 5.671\times10^4 z^{-1} + 3.18\times10^4 z^{-2} \\ -4176z^{-3} - 582.9z^{-4} + 43.69z^{-5} \end{pmatrix}$$

$$G_{5,w=0.5} = \begin{pmatrix} 1.022\times10^6 - 2.484\times10^6 z^{-1} + 2.048\times10^6 z^{-2} \\ -6.286\times10^5 z^{-3} + 4.033\times10^4 z^{-4} + 4033z^{-5} \\ \hline 30240 - 5.623\times10^4 z^{-1} + 3.097\times10^4 z^{-2} \\ -3759z^{-3} - 635.5z^{-4} + 41.68z^{-5} \end{pmatrix}$$

$$G_{5,w=0.6} = \begin{pmatrix} 1.024\times10^6 - 2.479\times10^6 z^{-1} + 2.032\times10^6 z^{-2} \\ -6.155\times10^5 z^{-3} + 3.678\times10^4 z^{-4} + 4236z^{-5} \\ \hline 30240 - 5.591\times10^4 z^{-1} + 3.043\times10^4 z^{-2} \\ -3489z^{-3} - 668.2z^{-4} + 40.19z^{-5} \end{pmatrix}$$

$$G_{5,w=0.7} = \begin{pmatrix} 1.025\times10^6 - 2.476\times10^6 z^{-1} + 2.02\times10^6 z^{-2} \\ -6.058\times10^5 z^{-3} + 3.416\times10^4 z^{-4} + 4382z^{-5} \\ \hline 30240 - 5.567\times10^4 z^{-1} + 3.003\times10^4 z^{-2} \\ -3291z^{-3} - 691.6z^{-4} + 39.01z^{-5} \end{pmatrix}$$

$$G_{5,w=0.8} = \begin{pmatrix} 1.026\times10^6 - 2.473\times10^6 z^{-1} + 2.011\times10^6 z^{-2} \\ -5.983\times10^5 z^{-3} + 3.216\times10^4 z^{-4} + 4491z^{-5} \\ \hline 30240 - 5.549\times10^4 z^{-1} + 2.972\times10^4 z^{-2} \\ -3139z^{-3} - 709.1z^{-4} + 38.04z^{-5} \end{pmatrix}$$

$$G_{5,w=0.9} = \begin{pmatrix} 1.026\times10^6 - 2.471\times10^6 z^{-1} + 2.005\times10^6 z^{-2} \\ -5.938\times10^5 z^{-3} + 3.095\times10^4 z^{-4} + 4556z^{-5} \\ \hline 30240 - 5.538\times10^4 z^{-1} + 2.953\times10^4 z^{-2} \\ -3047z^{-3} - 719.5z^{-4} + 37.44z^{-5} \end{pmatrix}$$

(14)

## B. Optimization of Chen-Vinagre Type Generating Function

For the realization of a semi-integrator with Chen-Vinagre type generating function (8) when employed via GA always produces the optimized parameter as $\alpha = 1$. This makes the realization restricted to Simpson only since it is a second order approximation with higher degree of accuracy in frequency domain. Optimization with Chen-Vinagre type of generating function (8) has also been carried out for 5$^{th}$ order IIR realization but no improvement in the frequency response has been found. The 3$^{rd}$ and 5$^{th}$ order optimum IIR filter with Chen-Vinagre type generating function is given by (15) and (16) respectively.

$$G_{3,opt} = \begin{pmatrix} 6.158\times10^6 + 3.266\times10^7 z^{-1} \\ +3.614\times10^7 z^{-2} - 9.263\times10^6 z^{-3} \\ \hline 3.373\times10^8 + 1.114\times10^9 z^{-1} \\ +8.847\times10^7 z^{-2} - 9.193\times10^8 z^{-3} \end{pmatrix} \quad (15)$$

$$G_{5,opt} = \begin{pmatrix} 3.771\times10^{23} + 2.967\times10^{24} z^{-1} \\ +6.819\times10^{24} z^{-2} + 2.453\times10^{24} z^{-3} \\ -4.675\times10^{24} z^{-4} - 5.358\times10^{23} z^{-5} \\ \hline 2.066\times10^{25} + 1.212\times10^{26} z^{-1} \\ +1.517\times10^{26} z^{-2} - 1.305\times10^{26} z^{-3} \\ -1.731\times10^{26} z^{-4} + 5.453\times10^{25} z^{-5} \end{pmatrix} \quad (16)$$

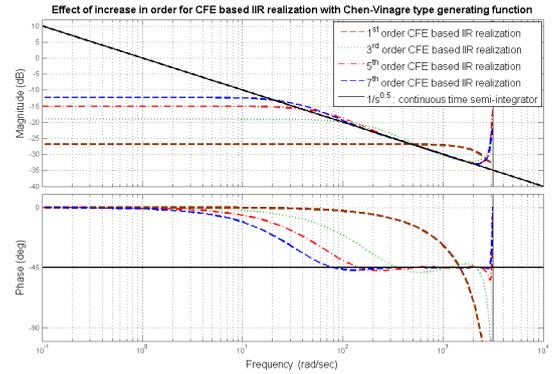

Fig. 6. Accuracies of different order of IIR realization of Al-Alaoui operator with Chen-Vinagre type generating function.

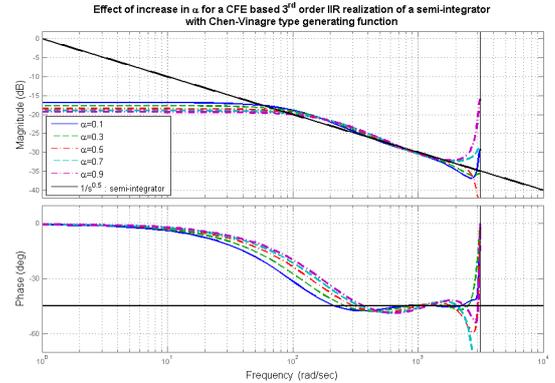

Fig. 7. Accuracies of different weights of Chen-Viangre type generating function interpolation for 3$^{rd}$ order IIR realization.

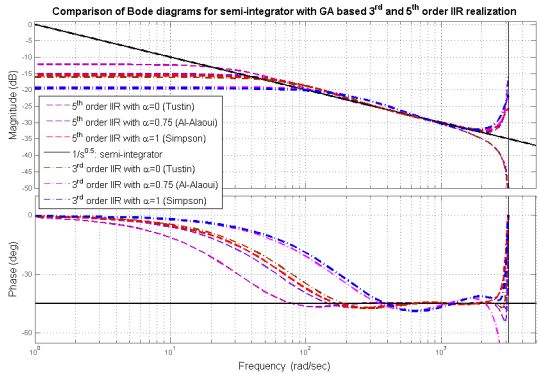

Fig. 8. Accuracies of GA based 3$^{rd}$ and 5$^{th}$ order IIR realization.

Fig. 6 and 7 presents effects of variation in the IIR filter order and weight ($\alpha$) of the Chen-Vinagre type generating function, similar to that presented in section IIB. The relative accuracies of Tustin, Simpson and Al-Alaoui operator based 3$^{rd}$ and 5$^{th}$ order IIR realization of a semi-integrator is also shown in Fig. 8.

IV. CONCLUSION

This paper shows that CFE based FO differentiator or integrator realizations can be improved using genetic algorithm. Proposed GA based IIR realization produces better accuracy while also keeping the order of realized filter low. Al-Alaoui's operator with GA based realization shows better accuracy compared to its original 3$^{rd}$ and 5$^{th}$ order realization for a FO semi-differentiator. GA based Chen-Vinagre type generating function reduces to the simple Simpson type generating function for 3$^{rd}$ and 5$^{th}$ order realization for a FO semi-integrator. Stability preserved discretaization of semi-differentiators via inversion of Chen-Vinagre type generating function by reflecting the unstable poles within unit circles has been detailed in [13]. Further investigation can be directed towards performance measure of the proposed optimization technique for varying differ-integrator orders i.e. without confining the study in semi-differintegrators only. The concept proposed in this paper can also be extended for higher order operators like [17], [18], [2] in future research.

REFERENCES

[1] Shantanu Das, "Functional fractional calculus for system identification and controls", Springer, Berlin, 2008.
[2] B.T. Krishna, "Studies on fractional order differentiators and integrators: a survey", *Signal Processing*, vol. 91, no. 3, pp. 386-426, March 2011.
[3] Guillermo E. Santamaria, Jose V. Valverde, Raquel Perez-Aloe and Blas M. Vinagre, "Microelectronic implementations of fractional-order integrodifferential operators", *ASME Journal of Computational and Nonlinear Dynamics*, vol. 3, no. 2, pp. 021301, April 2008.
[4] I. Podlubny, I. Petras, B.M. Vinagre, P. O'Leary, and L. Dorcak, "Analogue realization of fractional-order controllers", *Nonlinear Dynamics*, vol. 29, no. 1-4, pp. 281-296, 2002.
[5] A. Charef, "Analogue realization of fractional-order integrator, differentiator and fractional PI$^\lambda$D$^\mu$ controller", *IEE Proceedings-Control Theory & Applications*, vol. 153, no. 6, pp. 714-720, Nov 2006.
[6] Mohammad Adnan Al-Alaoui, "Discretization methods of fractional parallel PID controllers", *16$^{th}$ International Conf. on Electronics, Circuits, and Systems, ICECS 2009*, pp. 327-330, Dec. 2009.
[7] YangQuan Chen, Ivo Petras and Dingyu Xue, "Fractional order control-a tutorial", *American Control Conference, ACC '09*, pp. 1397-1411.
[8] YangQuan Chen, Blas M. Vinagre and Igor Podlubny, "Continued fraction expansion approaches to discretizing fractional order derivatives-an expository review", *Nonlinear Dynamics*, vol. 38, no. 1-4, pp. 155-170, 2004.
[9] D. Valerio and J. Sa da Costa, "Time-domain implementation of fractional order controllers", *IEE Proceedings-Control Theory and Applications*, vol. 152, no. 5, pp. 539-552, Sept. 2005.
[10] Blas M. Vinagre, YangQuan Chen, and Ivo Petras, "Two direct Tustin discretization methods for fractional-order differentiator/integrator", *Journal of the Franklin Institute*, vol. 340, no. 5, pp. 349-362, 2003.
[11] Chien-Cheng Tseng, "Design of FIR and IIR fractional order Simpson digital integrators", *Signal Processing*, vol. 87, no. 5, pp. 1045-1057, May 2007.
[12] YangQuan Chen and Kevin L. Moore, "Discretization schemes for fractional-order differentiators and integrators", *IEEE Transactions on Circuits and Systems-I: Fundamental Theory and Applications*, vol. 49, no. 3, pp. 363-367, March 2002.
[13] YangQuan Chen and Blas M. Vinagre, "A new IIR-type digital fractional order differentiator", *Signal Processing*, vol. 83, no. 11, pp. 2359-2365, Nov. 2003.
[14] Yang Zhu-Zhong and Zhou Ji-Liu, "An improved design for the IIR-type digital fractional order differential filter", *International Seminar on Future BioMedical Information Engineering, FBIE '08*, pp. 473-476, Dec. 2008, Wuhan, Hubei.
[15] G. Maione, "Continued fractions approximation of the impulse response of fractional-order dynamic systems", *IET Control Theory and Applications*, vol. 2, no. 7, pp. 564-572, July 2008.
[16] Richa Yadav and Maneesha Gupta, "Design of fractional order differentiators and integrators using indirect discretization approach", *2010 International Conference on Advances in Recent Technologies in Communication and Computing (ARTCom)*, pp. 126-130, Oct. 2010.
[17] Maneesha Gupta, Pragya Varshney and G.S. Visweswaran, "Digital fractional-order differentiator and integrator models based on first-order and higher order operators", *International Journal of Circuit Theory and Applications*, vol. 39, no. 5, pp. 461-174, May 2011.
[18] G.S. Visweswaran, P. Varshney, and M. Gupta, "New approach to realize fractional power in z-domain at low frequency", *IEEE Transactions on Circuits and Systems-II: Express Briefs*, vol. 58, no. 3, pp. 179-183, March 2011.
[19] M.A. Al-Alaoui, "Novel digital integrator and differentiator", *Electronics Letters*, vol. 29, no. 4, pp. 376-378, Feb. 1993.
[20] S.C. Ng, S.H. Leung, C.Y. Chung, A. Luk, W.H. Lau, "The genetic search approach: a new learning algorithm for adaptive IIR filtering", *IEEE Signal Processing Magazine*, vol. 13, no. 6, pp. 38-46, 1996.
[21] R. Storn, "Differential evolution design of an IIR-filter", *Proceedings of IEEE International Conference on Evolutionary Computation 1996*, pp. 268-273, May 1996, Nagoya, Japan.
[22] Swagatam Das and Amit Konar, "A swarm intelligence approach to the synthesis of two-dimensional IIR filters", *Engineering Applications of Artificial Intelligence*, vol. 20, no. 8, pp. 1086-1096, Dec. 2007.
[23] Adem Kalinli and Nurhan Karaboga, "Artificial immune algorithm for IIR filter design", *Engineering Applications of Artificial Intelligence*, vol. 18, no. 8, pp. 919-929, Dec. 2005.
[24] Nurhan Karaboga, "Digital IIR filter design using differential evolution algorithm", *EURASIP Journal on Applied Signal Processing*, vol. 8, no. 8, pp. 1269-1276, 2005.
[25] B. Luitel and G.K. Venayagamoorthy, "Differential evolution particle swarm optimization for digital filter design", *IEEE Congress on Evolutionary Computation, CEC 2008*, pp. 3954-3961, June 2008.